\documentclass[10pt,twocolumn]{article}

\usepackage[utf8]{inputenc}
\usepackage[english]{babel}
\usepackage{amsmath}
\usepackage{amsfonts}
\usepackage{amssymb}
\usepackage{mathpazo} 
\usepackage[scaled]{helvet}
\usepackage[T1]{fontenc}
\usepackage{url}
\usepackage{verbatim}
\usepackage{float}
\usepackage{caption}
\usepackage{subcaption}

\usepackage{graphicx}
\usepackage{xcolor}
\usepackage{booktabs}
\usepackage{algorithm}
\usepackage[noend]{algpseudocode}
\usepackage{changepage}

\usepackage[left=48pt,%
                right=42pt,%
                top=42pt,%
                bottom=60pt,%
                headheight=15pt,%
                headsep=10pt,%
                letterpaper,twoside]{geometry}%
                
\usepackage[labelfont={bf,sf},%
                labelsep=period,%
                figurename=Fig.,%
                singlelinecheck=off,%
                justification=RaggedRight]{caption}
                
\usepackage[numbers,sort&compress]{natbib}
\title{Subpicosecond dynamics of pre-plasma on a solid, formed by a ultra-high contrast, relativistic intensity pulse}

\usepackage{authblk}
\author[1,*]{Ankit Dulat}
\author[1]{C. Aparajit}
\author[1]{Anandam Choudhary}
\author[1]{Amit D. Lad}
\author[1]{Yash M. Ved}
\author[1,*]{G.Ravindra Kumar}

\affil[1]{Tata Institute of Fundamental Research, 1 Homi Bhabha Road, Colaba, Mumbai 400 005, India}
\affil[*]{Corresponding author: ankit$\_$090@tifr.res.in; and grk@tifr.res.in}

\date{}

\begin{document}

\twocolumn[
  \begin{@twocolumnfalse} 
    
    \maketitle
    
    \begin{abstract}
    Using spectral interferometry technique, we measured subpicosecond time-resolved pre-plasma scale lengths and early expansion ($<$ 12 ps) of the plasma produced by a high intensity (2$\times$10$^{18}$ W/cm$^{2}$) pulse with ultra-high contrast (10$^{-9}$). We measured pre-plasma scale lengths in the range of 3-15 nm. This measurement plays a crucial role in understanding the mechanism of laser coupling its energy to hot electrons and hence important for laser-driven ion acceleration and fast ignition approach to fusion.
    \vspace{2mm}
    \end{abstract}
    
  \end{@twocolumnfalse}
]

\section{Introduction}

With recent advances in lasers, the interaction of high-power femtosecond pulses of ultrahigh contrast (UHC), with solid density plasma is now accessible, facilitating high energy density science. Unlike pulses having intense picosecond wing and nanosecond pedestal where interaction occurs in expanding, lower than solid density plasma, UHC pulses cause ‘instantaneous’ excitation of solid density matter \cite{MURNANE531}.

The interaction mechanism of such clean pulses depends on well-defined initial conditions, especially on the scale length of the steep density (l$_s<<\lambda$)  pre-plasma \cite{Blanc1996,Landen1989} formed before the arrival of the main pulse. Earlier studies have shown, the critical dependence of pre-plasma scale length on the laser absorption \cite{Ping}, the energy distribution of heated electrons \cite{Macphee2010,Peebles2017}, angular distribution \cite{Santala2000}, and divergence of fast electron beam \cite{Ovchinnikov2013}. High harmonic generation from the interaction of ultra-intense pulses with solid density plasma also depends on the pre-plasma scale length \cite{Kahaly}. Various other applications such as the fast ignition approach to fusion \cite{Macphee2010} and laser-driven ion acceleration \cite{Gizzi2021} depend on the efficiency of coupling laser energy into fast electrons, which is sensitive to the scale length of preformed plasma. Therefore it becomes important to accurately measure the scale-length of the pre-plasma, which is extremely hard to measure and remains mostly unknown for high contrast relativistic pulses.
 
Until now, different techniques have been used to study the early evolution (after the arrival of the main pulse) of plasma with sub-picosecond time scale, but only a few have been used to study the pre-plasma conditions, which have their own limitations. Schlieren imaging technique \cite{Benattar1992} and shadowgraphy technique \cite{Dey2016} have been widely used to locate the critical density layer, however diffraction effects limit the spatial resolution to the order of incident wavelength. Spatial interferometry \cite{Adumi2004} has been used earlier, to measure the time evolution of density profiles but again it suffers form limited spatial resolution. Other techniques like time-resolved spectroscopy of back-reflected light and Doppler spectrometry \cite{Jana2019} are limited to observation of the dynamics at later delays because there is not much expansion/motion before the arrival of main pulse. Resonance absorption spectroscopy (RAS) \cite{Kieffer1989,Landen1989} has been used earlier, to measure the pre-plasma scale length by measuring plasma reflectivity differences between S and P-polarized light as a function of the angle of incidence, but it depends critically on the intensity and accurate modeling of the preformed plasma conditions, especially at high intensities. Unlike RAS, which is an indirect way of measuring pre-plasma evolution, frequency domain interferometry [16] (FDI or spectral interferometry) technique has an advantage of directly inferring the plasma dynamics from the measured phase shifts. Also, it can measure pre-plasma length scales with nanometer spatial resolution. It has been used earlier to measure the dynamics at positive delays of the plasma at relatively low intensities (<10$^{16}$ W/cm$^{2}$) \cite{Blanc1996,Geindre2001}.

In this paper, using frequency domain interferometry, we measured the scale length of the pre-plasma with nanometer spatial resolution and early expansion of the plasma at positive delays, created by 800 nm, high contrast (10$^ {-9}$), 27 femtosecond (fs) pulses at a peak intensity of 2$\times$10$^{18}$ W/cm$^{2}$.
\begin{figure*}[!ht]
    \begin{subfigure}{.55\linewidth}
        \centering
		\includegraphics[width=\columnwidth]{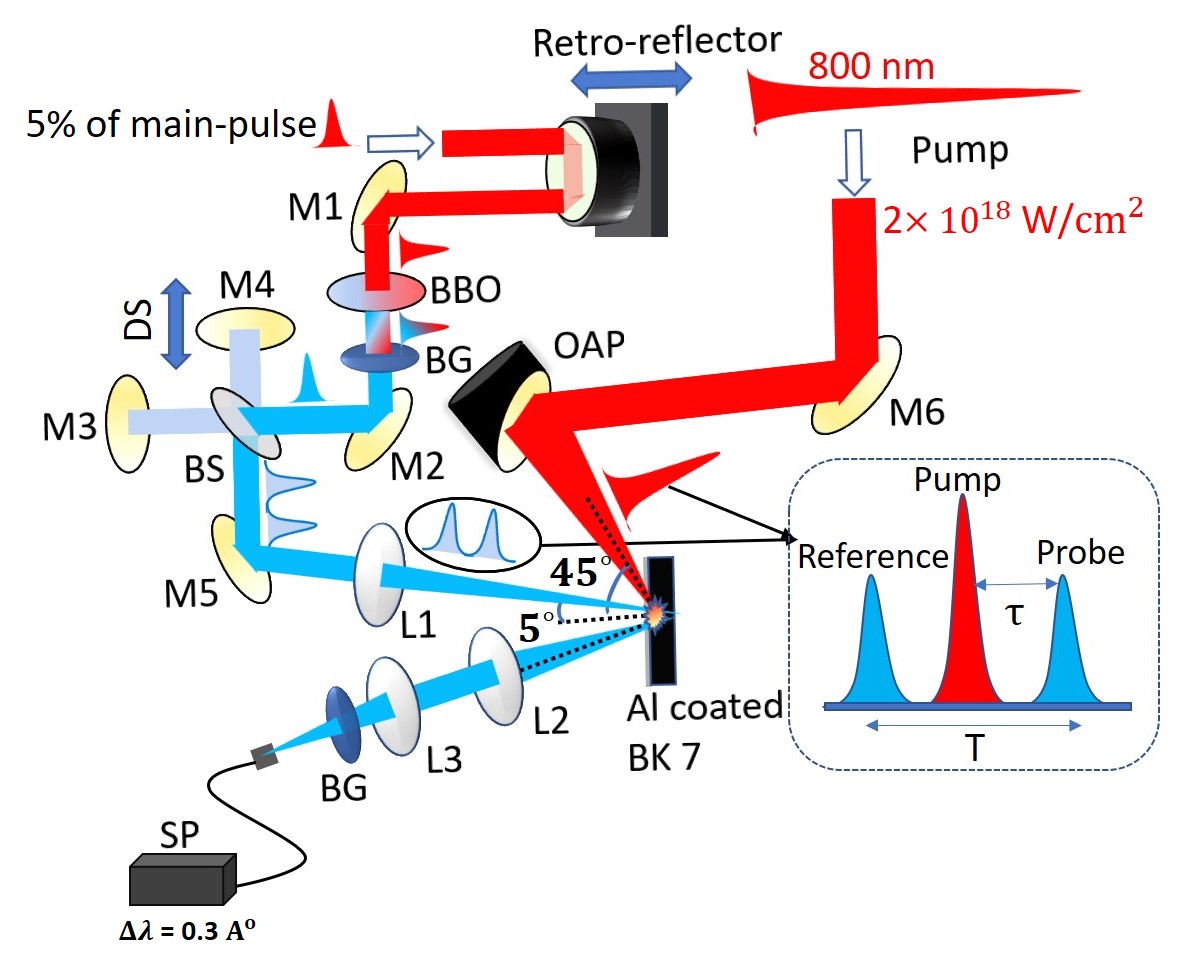}
		\caption{}
		\label{fig:setup}
    \end{subfigure}
    \begin{subfigure}{.45\linewidth}
        \centering
        \includegraphics[width = \columnwidth]{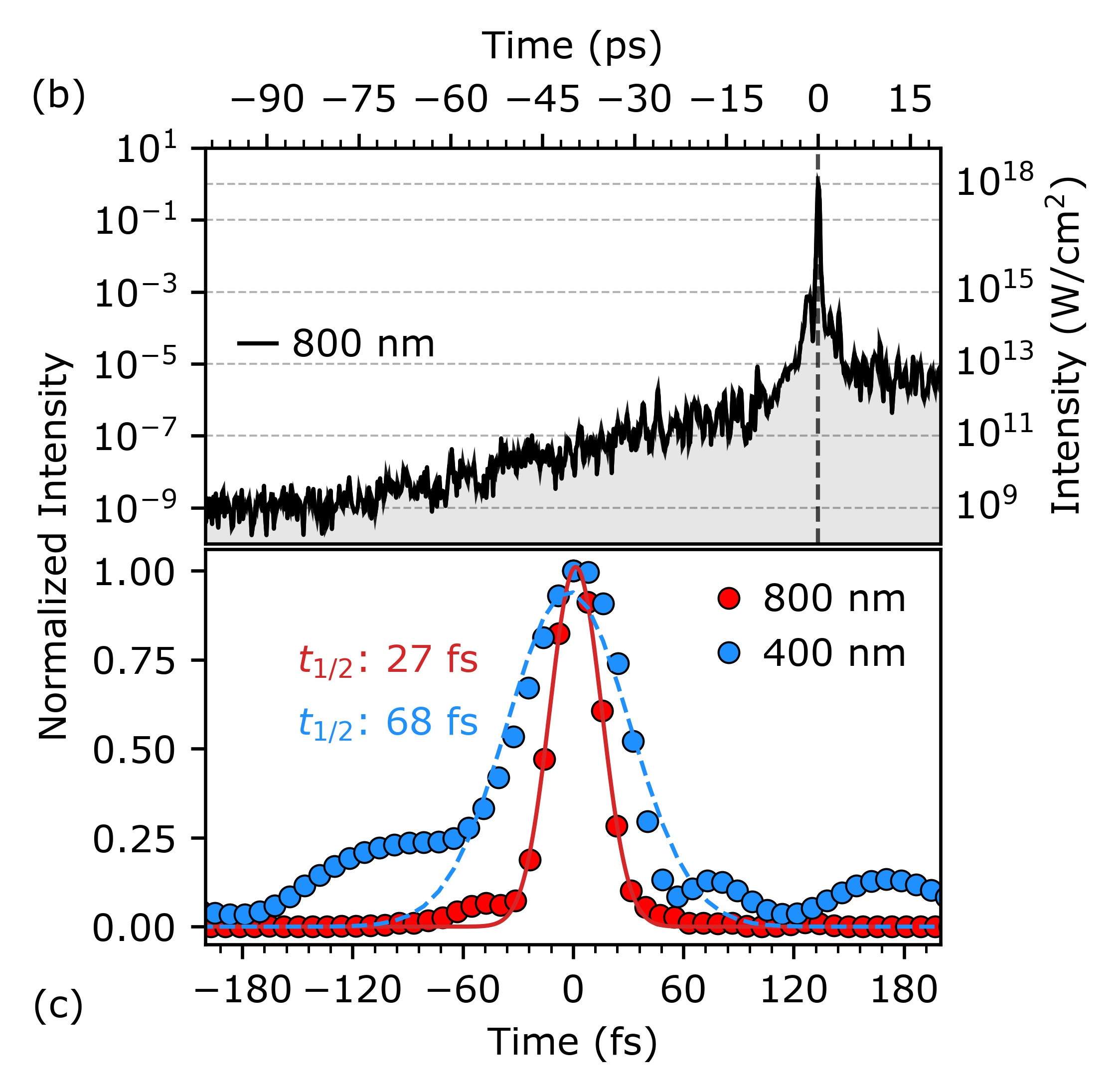}
        \label{fig:pulse_spec}
    \end{subfigure}
    \caption{(a) Schematic of experimental set-up for spectral interferometry. M1-M7: Mirrors, BBO: $\beta$-barium borate, BG: Blue-green (BG-39) filter, DS: Delay stage, BS: Beam splitter, L1-L3: Lenses, OAP: Off-axis parabola, SP: Spectrometer (b) Intensity contrast of 800 nm, measured using SEQUIOA \cite{Eckardt,ALBRECHT198159}. (c) Pulse-width of 800 nm (in red), measured using SPIDER \cite{Iaconis:98} and pulse-width of 400 nm (in blue), measured using SD-FROG \cite{Trebino-Kane}. FWHM of the two pulses is labeled by t$_{1/2}$. } 
    \label{fig:fig2}
\end{figure*}

\vspace{-4mm}
\section{Experiment}

 The experiment (Fig. \ref{fig:setup}) was performed with a 150-terawatt Ti:sapphire laser, which can deliver 800 nm, 27 femtosecond pulses. The measured picosecond intensity contrast and pulse width of the laser are shown in Fig. 1(b) and 1(c) respectively. 
 
P-polarized laser pulses (100 mJ pulse energy) were focused with an f/3 off-axis parabolic mirror (OAP) to a 7 $\mu$m spot at an incidence angle of 45$^{\circ}$, creating a peak intensity of 2$\times$ 10$^{18}$ W/cm$^{2}$. We used transparent dielectric target (BK7 glass) with Al coating of 200-nm thickness at the front. The target was moved between laser shots to provide a fresh surface to each laser shot. A small fraction (5$\%$) of the main laser beam was extracted and up-converted to its second harmonic (400 nm) by a $\beta$-barium borate (BBO) crystal (2 mm thick). The measured temporal profile of this second harmonic (SH) pulse is shown in Fig. 1(c). SH pulse was split into two pulses with a Michelson interferometer (MI): a reference pulse (arriving before the pump) and a probe pulse (reaching after the pump). A BG-39 filter was used after the BBO crystal to remove residual 800 nm light. One of the mirrors (M4) in MI is mounted on a high-precision motor stage which can be used to change the delay between the probe and reference pulses. Since the fringe spacing in FDI is inversely proportional to the delay \cite{Tokunaga1995} between these pulses, we have kept a fixed optimum delay (T) of 5.3 ps in our experiment (depending on the resolution of our spectrometer). The probe pulse was time-delayed with respect to the pump pulse using a high precision motorized retro-reflector delay stage.

The pair of two pulses (reference and probe) were focused onto the target at near-normal incidence ($\sim$5$^{\circ}$) with a convex lens (L1) to a spot of $\sim$120 $\mu$m (measured by knife-edge method).
The focal spot size of the probe and the reference beams was several times larger than the pump focal spot size to permit uniform illumination of the region under test.
Spatial overlap between the pump and the probe beams was ensured by high-resolution imaging. The reflected probe and reference pulses were collected by an f/2 plano-convex achromat lens and focused on single-mode fiber (Thorlabs, SM-300), which couples the light to a high resolution (0.3 \AA) spectrometer (OOI, HR-4000). The temporal overlap between the pump and the probe pulses was monitored by observing the probe reflectivity as a function of probe delay. The experiment was performed in a vacuum chamber at 10$^{-5}$ Torr.

\section{Results and Discussion}

\begin{figure}[t]
    \centering
    \includegraphics[width=0.95\linewidth]{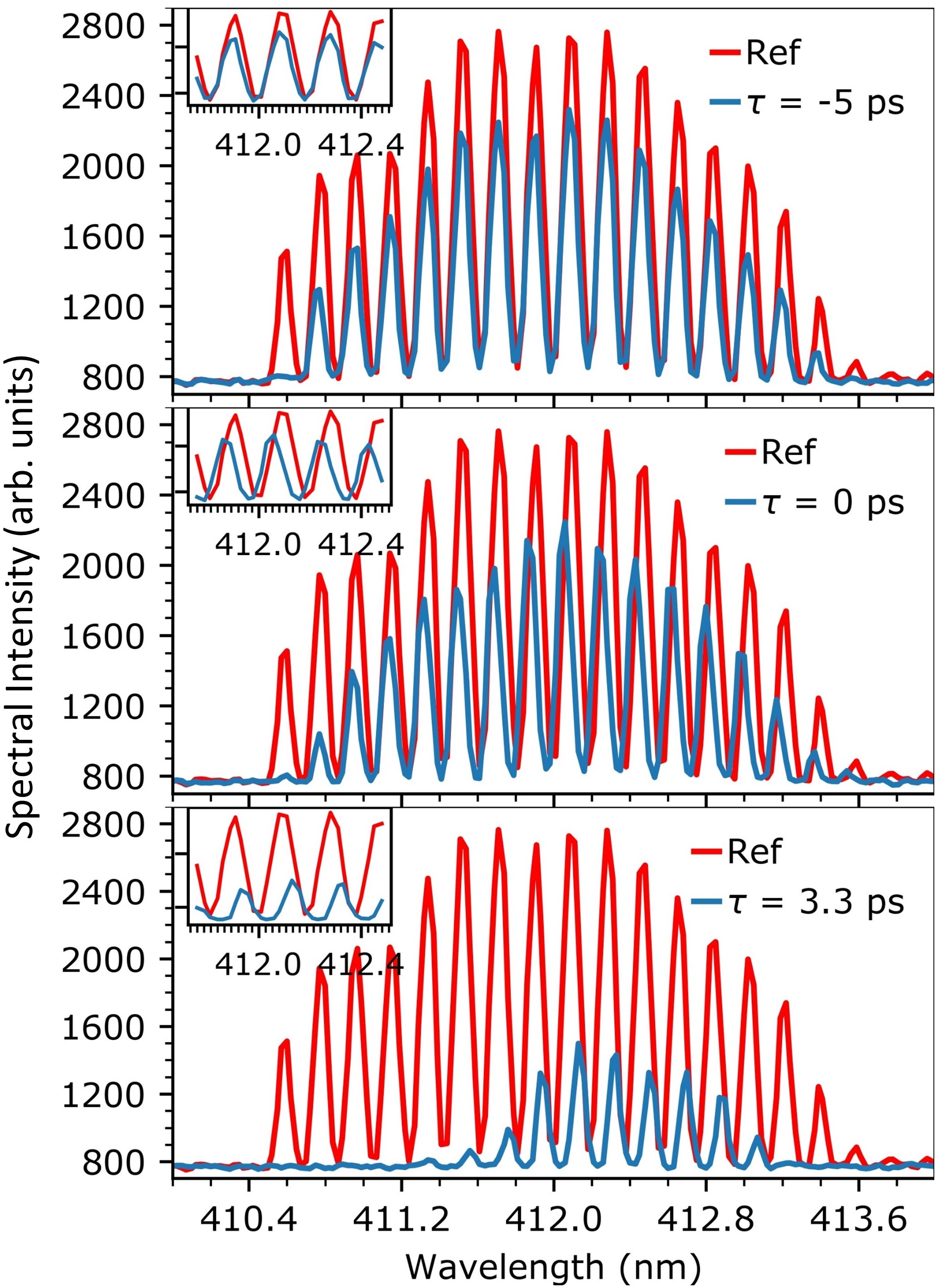}
    \caption{Measured interferogram at different time delays ($\tau$). Interferogram in red (Ref) is without plasma, in blue is with plasma on target. Insets show the zoomed central region of the interferogram.}
    \label{fig:raw_data}
\end{figure}

Spectral Interferometry relies upon measuring quantitative phase changes in probe pulse that is reflected from the plasma created by the pump laser. Fig. (\ref{fig:raw_data}) shows three spectral interferograms, each measured for a different delay ($\tau$) with respect to the pump. For each delay, the interferogram in ‘red’ is without any plasma on the target, while that in ‘blue’ is the phase-shifted one due to the interaction of probe pulse with the plasma. As observed, the interferogram gets more phase-shifted and its amplitude decreases as plasma evolves.
The phase shift in the probe pulse due to plasma has been calculated from the measured modulated spectrum using the Fast-Fourier-Transform (FFT) based numerical algorithm \cite{Takeda1982}. Details of the phase extraction procedure are given in the Appendix.

The two main possible contributions to this phase change are (1) changes in the refractive index of the material between the reflecting surface (critical density layer) and the observer, due to rapid ionization of the material and subsequent change in the electron density distribution. (2) due to the rapid expansion of the plasma and motion of the reflecting surface (Doppler phase shift).

\begin{figure}[H]
    \centering
    \includegraphics[width=\columnwidth]{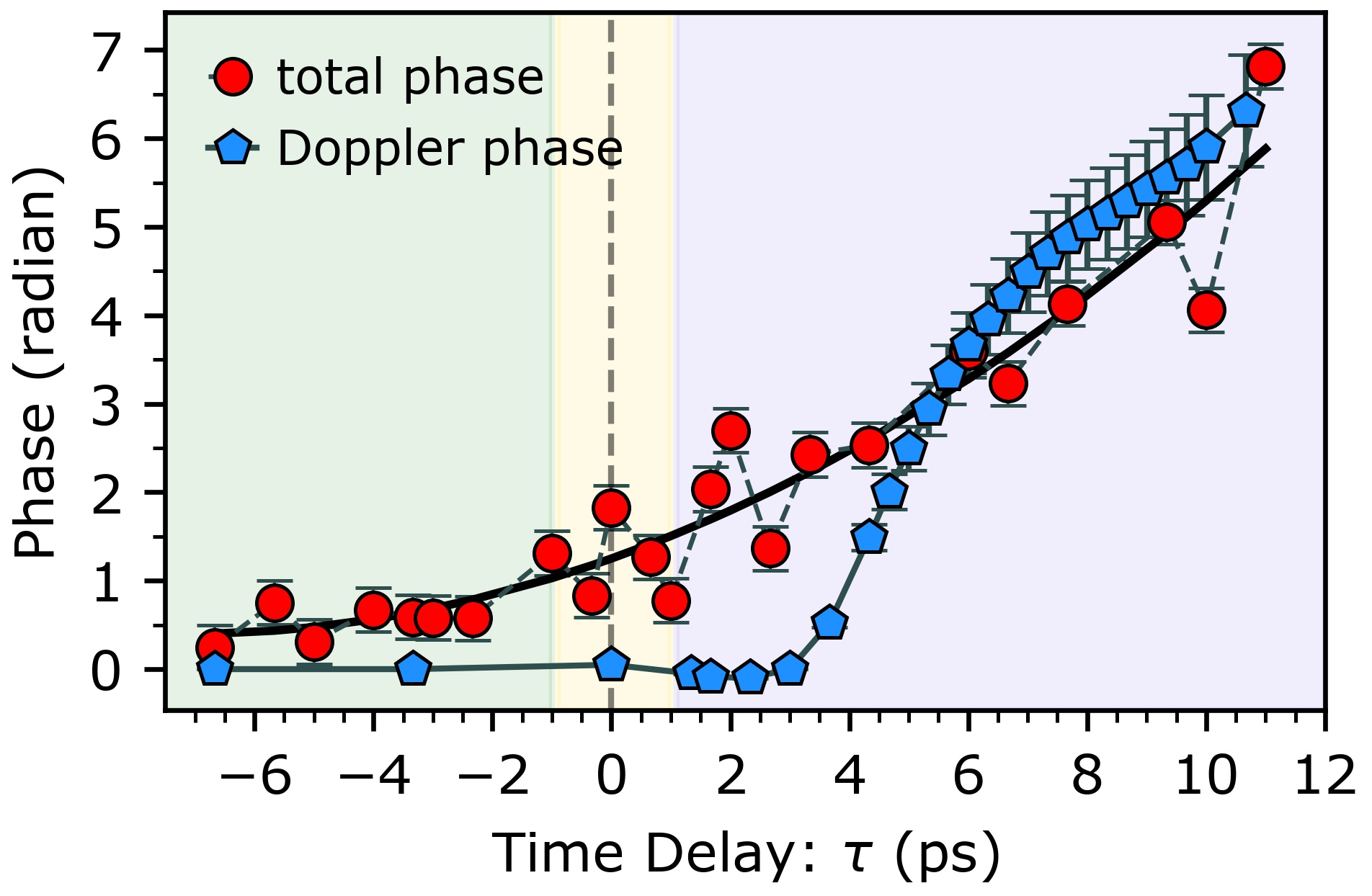}
    \caption{Time-resolved total phase shift calculated from FDI data, is compared with Doppler phase shift, calculated using the earlier Doppler spectrometry data (Jana et al. \cite{Jana2019}). }
    \label{fig:phase}
\end{figure}

The extracted total phase is shown in Fig. (\ref{fig:phase}). The error bars are due to shot-to-shot fluctuations of input laser energy and background vibrations which perturbs the interferogram. We observed a non-zero but small phase change at negative delays of -6.6 ps (before the main fs pump pulse hit the target) which is indicative of presence of significant plasma. Since our fs pulse doesn’t have any nanosecond prepulse, we expect this phase change to be mainly due to the ionization of the material by picosecond (ps) pedestal, which results in a very steep density profile. Subsequently, there is a tiny bit of gradual increase in this phase shift as we approach time-zero. Around time-zero ($\tau$ = 0), there is a rapid fluctuation in the phase. This phase shift is mainly due to the intense ionization of the target around the peak of the pulse and subsequent explosive expansion of low-density fast electron cloud (accelerated by various mechanisms) escaping the target. The oscillation in the phase around $\tau$ = 0 could be related to the ionization by intense (10$^{15}$ W/cm$^{2}$) pre-pulse (at $\sim$ -1 ps) and post-pulse (at $\tau$ < 5 ps ) as shown in Fig. 1(b).  

The phase shift at later time delays is mainly due to the rapid expansion of background plasma which has been heated through energy transfer from the fast electrons \cite{Jana2019}.  
 This phase shift can not be fitted with a linear curve, indicating that at early times (< 10 ps) plasma is not expanding with a constant velocity, which is consistent with earlier measurements (at similar experimental conditions) using Doppler spectrometry.

The total phase shift experienced by the probe beam can be expressed as  the path integral through the plasma up to point where the beam is reflected (z$_{c}$) and back ( z = 0 being the target surface intial position) \cite{Antici2007}.

\begin{equation} \label{eq:5}
\begin{split}
\phi(t) & = 2\int^{z=z_{c}(t)}_{\infty} \frac{\omega}{c}\sqrt{\epsilon(s)} ds \\
 & \approx \frac{2\omega}{c}\left (\int^{z=z_{c}(t)}_{\infty}ds - \frac{1}{2n_{c}}\int^{z=z_{c}(t)}_{\infty}n_{e}(s,t)ds\right)
\end{split}
\end{equation}

where s is the coordinate along beam path, $\omega$ is the angular frequency, c is the speed of light in vaccum, $\epsilon$ is the permittivity that can be approximated as $\epsilon$=1-(n$_{e}$/n$_{c}$) with n$_{e}$, n$_{c}$ is the local electron density and critical density respectively . The imaginary part in $\epsilon$ (due to ion-electorn collisions) has been neglected since it mainly causes absorption of the probe pulse, contributing weakly to the phase change. The measured phase shift is $\Delta \phi = \phi - \phi_{0}$, where 

\begin{equation}
    \phi_{0}(t) = 2\int^{0}_{\infty} \left(\omega/c\right) ds
\end{equation}
is the phase experienced by the reference pulse. The first term in total phase shift is only due to the motion (Doppler phase shift) of the reflecting surface and can be re-expressed as:
\begin{equation}
    d\phi/dt = 2\omega_{0} \left(nu/c\right) cos(\theta)
\end{equation}

where $u$ is the velocity of the critical surface and $n$ is the refractive index. Using the velocity data from earlier measurements of the Doppler spectrometry \cite{Jana2019} at similar experimental conditions (I = 4$\times$10$^{18}$ W/cm$^{2}$), we have calculated the contribution of Doppler phase in total phase shift (Fig. \ref{fig:phase}). As observed, for time-delays less than 4 ps, the phase shift is mainly due to the ionization effects and motion of fast electron cloud. It is only after 4 ps, contribution of doppler phase starts increasing and then becomes dominant at larger time delays. 

Assuming an exponential density profile \cite{Peebles2017} n$_{e}$=$\frac{n_{0}}{1+\exp{(z/l_{s})}}$, we calculated the length scale ($l_{s}$) of the
plasma (along target normal) from the measured phase shifts using Eq. (\ref{eq:5}). 
\begin{figure}[t]
    \centering
    \includegraphics[width=0.95\linewidth]{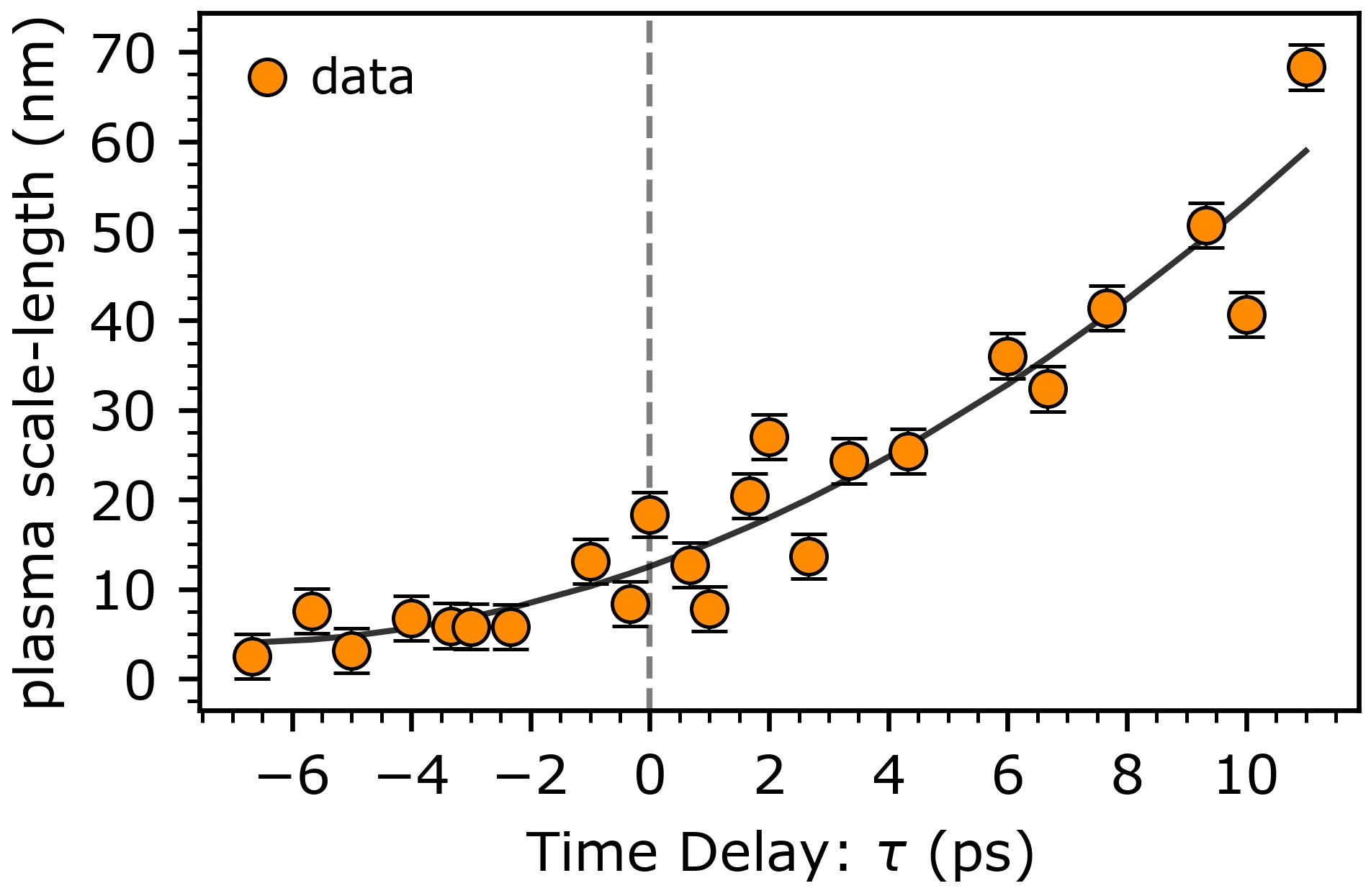}
    \caption{Calculated time-resolved plasma scale length}
    \label{fig:length_scale}
\end{figure}

\begin{figure}[H]
    \centering
    \includegraphics[width=\linewidth]{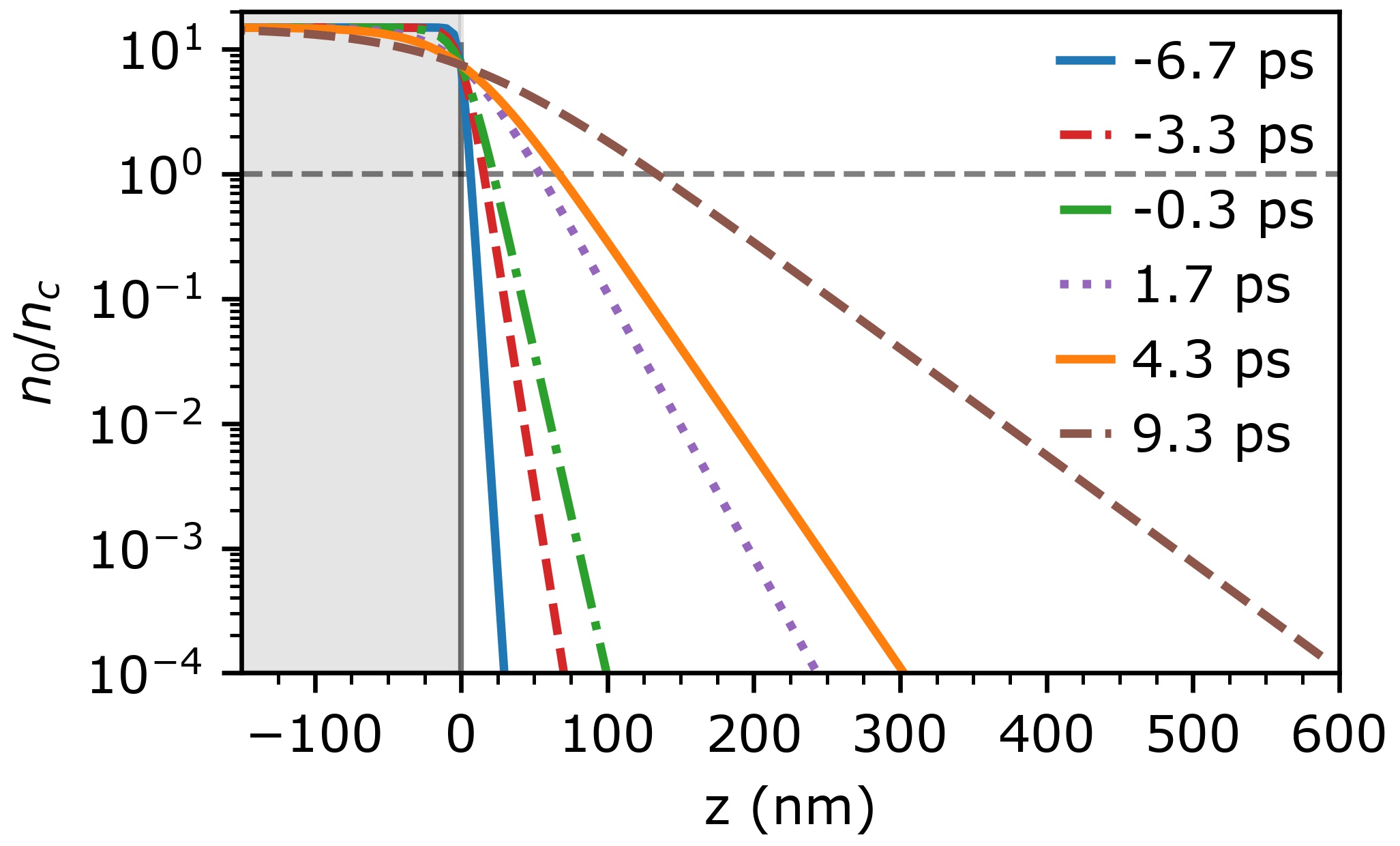}
    \caption{Time varying density profile of the plasma (Black dashed horizontal line indicates the critical density)}
    \label{fig:density_profile}
\end{figure}
Fig. (\ref{fig:length_scale}) shows the calculated length scales of the plasma, and corresponding density profiles at various time delays ($\tau$) are shown in Fig. (\ref{fig:density_profile}). From the measured phase shifts, we obtained the pre-plasma length scale of 15 nm ($\sim\lambda$/50) just before the main fs pulse hits the target.  Measured length scale is consistent with earlier results of hydrodynamic simulation \cite{Cerchez2018} with similar experimental conditions (same target, pump pulse with same contrast). 
Also, the total phase shift at $\tau$ > 5 ps is consistent with the Doppler phase shift calculated with earlier experimental data of Doppler spectrometry \cite{Jana2019}.

\section{Conclusions}
We have investigated the dynamics of solid density pre-plasma created by ultra high-contrast intense laser pulses using spectral interferometry. We measured pre-plasma scale length with nanometer spatial resolution and on the sub-ps time scale. This technique is very sensitive and can measure even the tiniest phase changes occurring due to the ionization. So, by appropriate modeling/simulations of pre-plasma conditions, it, therefore, is possible to map out the picosecond pedestal of ultra-high contrast, high-intensity pulses by measuring the phase changes at times before the main pulse hits (negative delays). This method may offer an advantage over other non-linear techniques \cite{Eckardt,ALBRECHT198159} used to measure ps-contrast, which are sensitive to the wavelength and struggle to measure contrast better than 10$^{-10}$.\\

\noindent\begin{Large}
\textbf{Funding}.\end{Large} GRK acknowledges partial support from J.C. Bose Fellowship grant (JBR/2020/000039) from the Science and Engineering Board (SERB), Government of India. \\

\noindent\begin{Large}
\textbf{Acknowledgement}\end{Large}
We thank Sunil B Shetye for his help in setting up vibration-free interferometer.
 \\


\bibliography{main}
\bibliographystyle{ieeetr}


\vspace{10mm}
\noindent\begin{Large}
\textbf{Appendix} \end{Large} \\

The interference signal with pump excitation for each pump-probe delay ($\tau$) can be expressed as \cite{Tokunaga1995}:
\begin{equation} \label{eq:1}
    I'(\omega) = |E_{pr}(\omega)|^{2}\left(1+R+2\sqrt{R}\cos(\omega T - \Delta\phi(\omega))\right)
\end{equation}
where  $\Delta\phi$ = n$_{r}\omega$L/c is the phase change due to the plasma, $R = \exp\left(-2n_{\mathrm{img}}\omega L/c \right)$ is the reflection cofficient, L is the length scale of the plasma and n$_{r}$, $n_{\mathrm{img}}$ are real and imaginary part of refractive index of plasma respectively. \\
To extract the phase change ($\Delta\phi$), we take the Fast-Fourier transform (FFT) of the measured spectrum (\ref{eq:1}), which gives the peaks centred at 0, $\pm T$  which is given by expression

\begin{equation} \label{eq:2}
\begin{split}
F[I'(\omega)](t')  = &\left( 1 + R \right)G(t') + \sqrt{R}\exp\left(i \Delta\phi \right)G\left(t' - T\right) \\
& + \sqrt{R}\exp\left(-i \Delta\phi \right)G\left(t' + T\right)
\end{split}
\end{equation}
where $G(t')$ is the Fourier transform of |E$_{pr}(\omega)|^{2}$. Each of the two peaks at $t' = \pm T$ contains the information about phase change ($\Delta\phi$) as well as reflection coefficient ($R$). So we filter out right side peak ($t' = T$) and do the inverse FFT to get 
\begin{equation} \label{eq:3}
    I'(\omega) = |E_{pr}(\omega)|^{2}\sqrt{R}\exp\left(i(\omega T + \Delta\phi)\right)
\end{equation}
A similar analysis of reference interferogram (without pump excitation) will give
\begin{equation} \label{eq:4}
    I(\omega) = |E_{pr}(\omega)|^{2}\exp\left(i\omega T\right)
\end{equation}
Now by just dividing equation (\ref{eq:3}) by equation (\ref{eq:4}), we get: P($\omega$) = $\sqrt{R}\exp(i\Delta\phi)$. \\

The imaginary part of logarithm of P($\omega$) gives the required phase shift ($\Delta\phi$). The phase so obtained is indeterminate to a factor of 2$\pi$. Also, a computer-generated function subroutine gives a principal value ranging from -$\pi$ to $\pi$. These discontinuities in the phase can be corrected by a simple algorithm of phase unwrapping \cite{Takeda1982}.

\end{document}